# Anomalous Hall effect in Weyl semimetal half Heusler compounds RPtBi (R = Gd and Nd)

Chandra Shekhar [a], Nitesh Kumar [a], V. Grinenko [b,c], Sanjay Singh [a], R. Sarkar [b], H. Luetkens [d], Shu-Chun Wu [a], Yang Zhang [a], Alexander C. Komarek [a], Erik Kampert [e], Yurii Skourski [e], Jochen Wosnitza [b,e], Walter Schnelle [a], Alix McCollam [f], Uli Zeitler [f], Jürgen Kübler [g], Binghai Yan [a], H.-H. Klauss [b], S. S. P. Parkin [h], C. Felser [a]

[a] *Max Planck Institute for Chemical Physics of Solids, 01187 Dresden, Germany*
[b] *Institute for Solid State and Materials Physics, TU Dresden, 01069 Dresden, Germany*
[c] *IFW Dresden, 01069 Dresden, Germany*
[d] *Laboratory for Muon Spin Spectroscopy, Paul Scherrer Institute, CH-5232 Villigen PSI, Switzerland*
[e] *Dresden High Magnetic Field Laboratory (HLD-EMFL), Helmholtz-Zentrum Dresden-Rossendorf, 01328 Dresden, Germany*
[f] *High Field Magnet Laboratory (HFML - EMFL), Radboud University, 6525 ED, Nijmegen, The Netherlands.*
[g] *Institute for Solid State Physics, TU Darmstadt, 64289 Darmstadt, Germany*
[h] *Max Planck Institute of Microstructure Physics, 06120 Halle, Germany*

**Abstract**

Topological materials ranging from topological insulators to Weyl and Dirac semimetals form one of the most exciting current fields in condensed-matter research. Many half-Heusler compounds, RPtBi (R= rare earth) have been theoretically predicted to be topological semimetals. Among various topological attributes envisaged in RPtBi, topological surface states, chiral anomaly and planar Hall effect have been observed experimentally. Here, we report on an unusual intrinsic anomalous Hall effect (AHE) in the antiferromagnetic Heusler Weyl semimetal compounds GdPtBi and NdPtBi that is observed over a wide temperature range. In particular, GdPtBi exhibits an anomalous Hall conductivity of up to 60 $\Omega^{-1}cm^{-1}$ and an anomalous Hall angle as large as 23%. Muon spin resonance (μSR) studies of GdPtBi indicate a sharp antiferromagnetic transition ($T_N$) at 9 K without any noticeable magnetic correlations above $T_N$. Our studies indicate that Weyl points in these half-Heuslers are induced by a magnetic field via exchange-splitting of the electronic bands at or near to the Fermi energy which is the source of the chiral anomaly and the AHE.

## Introduction

Topological materials are a family of quantum materials that are of much interest today as they possess a range of novel phenomena which are promising for technological applications. Among these materials Weyl semimetals (WSM) are three-dimensional (3D) analogues of graphene, in which the conduction and valence bands disperse linearly through nodes, the Weyl points, in momentum space (1). Weyl points are singular monopoles of the Berry curvature, an intrinsic property of the electron's wave function, with "+" or "–" chirality, and, thus, always come in pairs. A hallmark of a WSM is the existence of Fermi-arc surface states (1), which connect each pair of Weyl nodes. The recent discovery of time-reversal-invariant WSMs in the TaAs-type transition-metal monopnictides (2, 3) was demonstrated via the observation of Fermi arcs (4, 5). Another hallmark of a WSM is the so-called chiral anomaly which arises from topological charge pumping between Weyl points within a pair. This gives rise to several unusual properties including a longitudinal negative magnetoresistance (nMR) (6). Furthermore, time-reversal breaking WSMs are anticipated to exhibit an anomalous Hall effect (AHE) due to the net Berry flux that is proportional to the separation of the Weyl points that have opposite chirality (7). Among the large family of WSMs, $ZrTe_5$ has recently been identified as exhibiting an AHE (8), whose magnitude is sensitive to the position of Weyl points relative to the Fermi energy ($E_F$).

The extended family of Heusler compounds provides a unique platform since these materials possess a wide range of tunable structural and physical functionalities, ranging from topological insulators (9-11), to unconventional superconductivity (12). Topological surface states in several Heusler compounds have recently been observed by angle-resolved photoemission studies (13). Among the family of half-Heusler compounds, the Weyl semimetal GdPtBi is significant because it exhibits a chiral anomaly (14) , a planar Hall effect (15), a linear optical conductivity (16) and triple-point fermions (17) close to the $E_F$.

In this paper, we consider magnetic lanthanide half-Heusler compounds formed from RPtBi, where R is a lanthanide or Y. These compounds have an asymmetric inversion crystal structure (space group No. 216, $F\bar{4}3m$). RPtBi is composed of three interpenetrating *fcc* lattices (Fig. 1A) so that along the [111] direction, the structure can be described as a metallic multilayer formed from successive atomic layers of rare-earth, platinum and bismuth. GdPtBi (18, 19) as well as NdPtBi (20) are antiferromagnetic (AFM) metals at low temperatures below their corresponding Néel temperatures, $T_N =$ 9.0 K and 2.1 K, respectively. The Gd spins order antiferromagnetically without any canting in zero magnetic field (as evidenced from μSR, discussed later) and saturate into a fully spin-aligned magnetic state in high magnetic fields at low temperatures (e.g. 25 T at 1.4 K (Fig. 1B)). This is in contrast to a previous report where the AHE between 3 and 5 T is attributed to a spin texture Berry phase (21).

The calculated band structures of GdPtBi are given in Fig. S1. A simplified schematic version in Fig. 1C shows how the exchange field affects the band splitting, forming Weyl points; the band inversion between the $\Gamma_8$ and $\Gamma_6$ bands results in a gapless semimetal with degenerate $\Gamma_8$ bands at the $E_F$ for all lanthanide RPtBi compounds (9). When R = Y, the compound is non-magnetic and we do not expect such band splitting but most of the other lanthanides gives rise to magnetism. For example, GdPtBi and NdPtBi exhibit magnetism arising from their 4*f* electrons preserving the $\Gamma_8$ - $\Gamma_6$ band inversion. The magnetic structures of these compounds are different: GdPtBi is a type-II antiferromagnet (18, 19) whereas the magnetic structure of NdPtBi is of type-I (20). Additionally, one should note that the magnetic moments and *f*-level fillings are distinct for neodymium and gadolinium and the exchange field is sufficiently large to reveal, for example, four pairs of Weyl points that appear in GdPtBi when $B \parallel [111]$ (Fig. 1D). Interestingly, a nMR was reported for GdPtBi (14) when a magnetic field, *B* was applied parallel to the current direction. This was attributed to a chiral anomaly associated with a WSM. However, the existence of Weyl points was attributed to an external magnetic field induced Zeeman splitting.

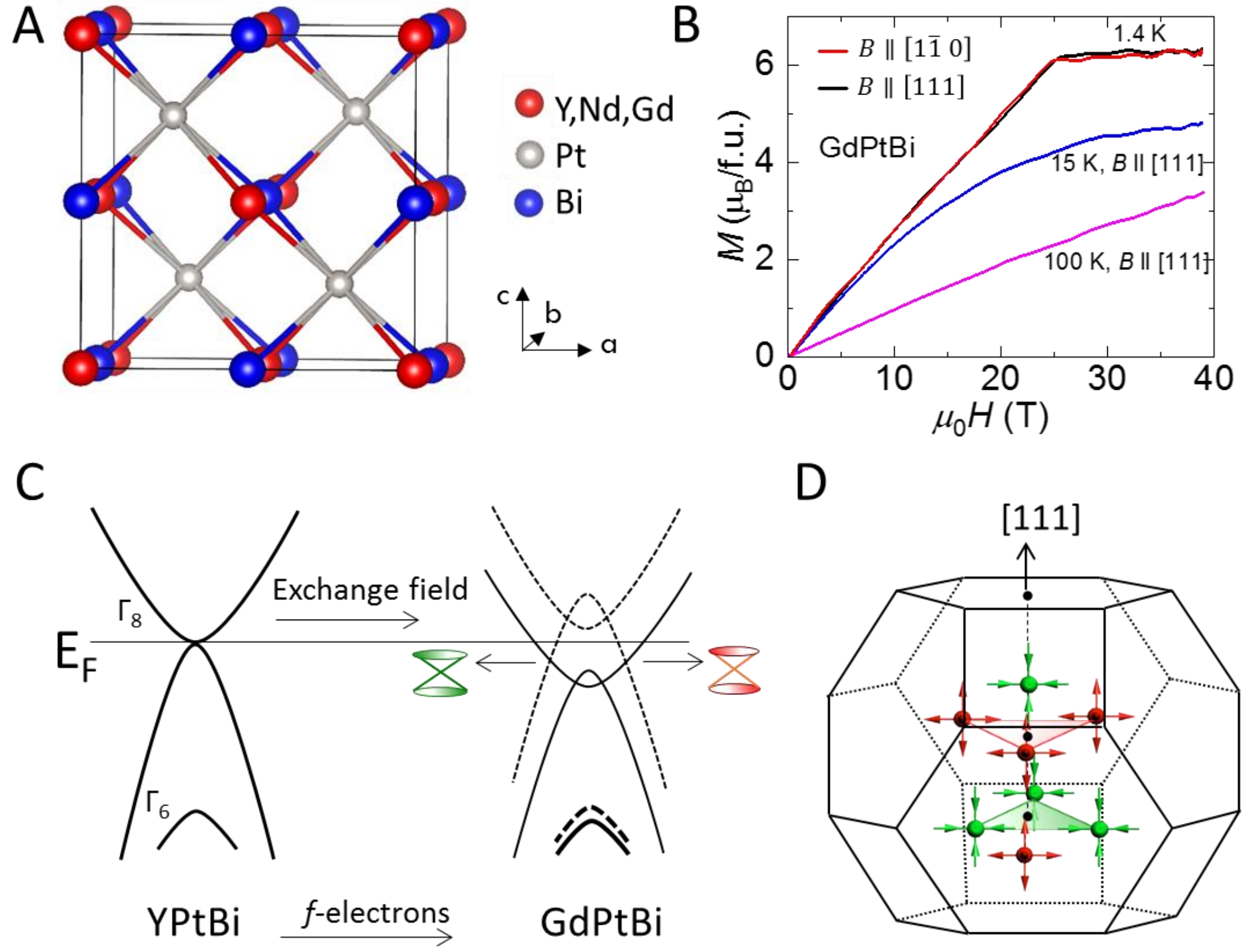

*Fig. 1. Structure, magnetization and evolution of Weyl points. (A) Structure of cubic unit cell of RPtBi (R=Y, Gd or Nd), (B) Magnetization in pulsed magnetic fields up to 40 T for GdPtBi at different temperature for B parallel to [111]of and [1$\bar{1}$0]. At T = 1.4 K, the saturation magnetization is ~25 T with a magnetic moment per Gd of almost 6.5 $\mu_B$ and magnetization is independent of crystallographic directions. (C) Schematic comparison of the calculated band structures of YPtBi and GdPtBi. The exchange field from the rare earth elements moments lifts the spin-degeneracy of the $\Gamma_8$ and $\Gamma_6$ bands and induces Weyl points: the green and red hourglasses represent Weyl cones with opposite chirality. (D) Distribution of Weyl points in the first Brillouin zone of GdPtBi when the magnetic moments are along [111] taken as an example. Green and red spheres represent "–" and "+" chirality, and the arrows are the Berry curvature vectors.*

The crystals studied here have been extensively investigated, as shown by x-ray diffraction, temperature-dependent resistivity, charge-carrier density, mobility, and specific heat (see Figs. S5 – S7). Both GdPtBi and NdPtBi exhibit antiferromagnetic-semi-metallic behaviors with low charge-carrier densities and high mobilities, similar to that reported in (22). We have carried out extensive measurements of the field-dependent Hall resistivity, $\rho_{xy}$, and longitudinal resistivity, $\rho_{xx}$ at various temperatures for the well-oriented crystal GdPtBi-1. At a constant value of $\theta = 0^o$ ($I \perp B$), where $\theta$ is the angle between the magnetic field and current directions, we find that $\rho_{yx}$ reveals a small hump (marked region) around 2 T (Fig. 2A) at low temperatures. In $\rho_{xx}$, this anomaly is reflected as a pronounced dip at the same magnetic field, as can be clearly seen in Fig. 2B. This feature shifts to slightly higher magnetic fields as the temperature is increased. These anomalies in $\rho_{yx}$ and $\rho_{xx}$ almost disappear for $T \geq 60$ K. Consequently, the calculated Hall conductivity, $\sigma_{xy}$, (Fig. 2C) also exhibits a peak in the same *H*-*T* regime. This anomaly in $\rho_{yx}$ is attributed to the anomalous Hall effect (AHE) which is usually observed in ferromagnetic materials (23). Recently, it has been shown that non-collinear antiferromagnets with zero net magnetization can produce a large AHE when their electronic structure exhibits a non-vanishing Berry-curvature (that acts like a large fictitious magnetic field) (24-27). Zero-field (ZF) μSR data are consistent with collinear antiferromagnetic order below $T_N \sim 9$ K (see below). The AHE persists well above $T_N$, where the presence of static magnetic correlations is excluded by our μSR experiments.

Theoretical studies have predicted that a large AHE can also be realized in Weyl semimetals that exhibit a chiral anomaly, where the anomalous Hall conductance (AHC) is proportional to the separation between the Weyl nodes (7). As discussed, GdPtBi is an ideal Weyl semimetal with field-induced Weyl nodes near the $E_F$ (Fig. 1) and therefore exhibits a nMR when $\theta = 90^o$ ($I \parallel B$) (Fig. S9). The AHC for GdPtBi-1 and NdPtBi at 2 K are 60 $\Omega^{-1}$cm$^{-1}$ and 14 $\Omega^{-1}$cm$^{-1}$, respectively. Similar results were obtained on a second crystal GdPtBi-2. Our findings clearly suggest a major role of the Weyl points in the AHE in GdPtBi and NdPtBi.

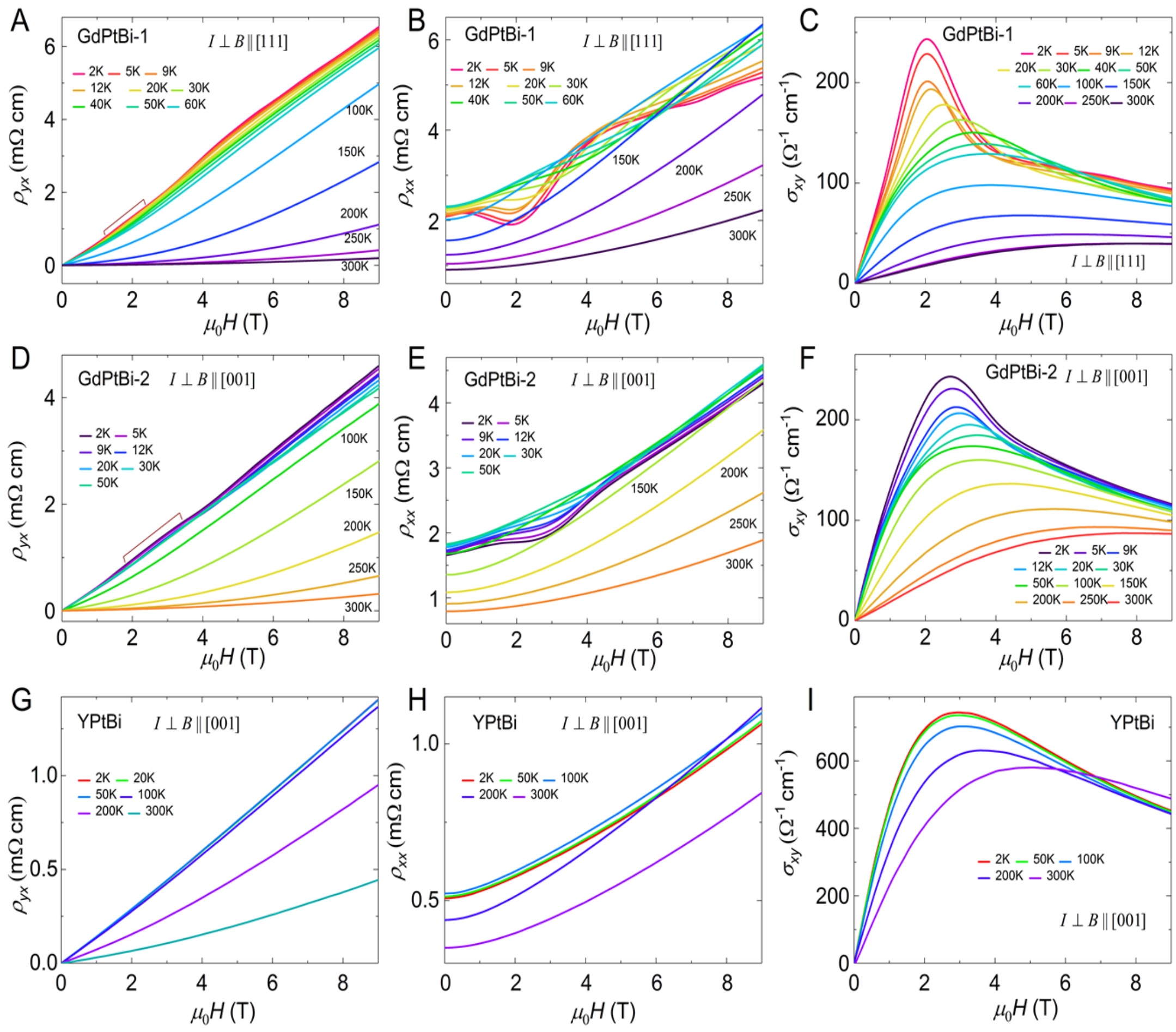

*Fig. 2. Field dependence of resistivity at various temperatures. (A), (D), (G) Hall resistivity ($\rho_{yx}$); (B), (E), (H) resistivity ($\rho_{xx}$) and (C), (F), (I) corresponding derived Hall conductivities ($\sigma_{xy}$). Upper two and lower panels are for GdPtBi and YPtBi compounds, respectively where $\sigma_{xy}$ is calculated from the relation* $\sigma_{xy} = \frac{\rho_{yx}}{\rho_{yx}^2+\rho_{xx}^2}$.

To investigate the AFM ordered state at low temperatures from the local point of view and to examine the presence of magnetic correlations in the paramagnetic state, we have performed in the time domain of the ZF spectra as shown in the supplementary material (Fig. S14). This demonstrates a well-defined static internal field at the muon site due to the static ordered moment of Gd spins. The frequency of the oscillations is given by the internal field at the muon stopping site, which is related to the AFM order parameter. The temperature dependence of the internal field follows the predictions for a conventional Heisenberg AFM state with a spin-wave ZF-μSR experiments on GdPtBi single crystals. Below $T_N \sim 9$ K, we observe clear oscillations contribution at low temperatures (Fig. 3A) (28). The amplitude of the oscillations is isotropic for two orthogonal components of the muon spin polarization (inset of Fig. 3A). The absolute value of the amplitude is about 60% of the total asymmetry, indicating that the internal fields at the muon stopping site have an angle of about 50° with respect to the muon spin polarization. This is consistent with the AFM structure proposed previously (18, 19). The muon spin depolarization rate $\lambda_L$ (obtained from the fit of the ZF-μSR time spectra as shown in Fig. S14B), which is a measure of the fluctuations/correlations of the spins, does not show any noticeable divergent behavior across $T_N$ (Fig. 3B). Further, in the paramagnetic state above $T_N$, $\lambda_L$ is temperature independent and isotropic for two orthogonal components of the muon spin polarization. The overall behavior suggests the absence of any significant magnetic correlations in GdPtBi above $T_N$.

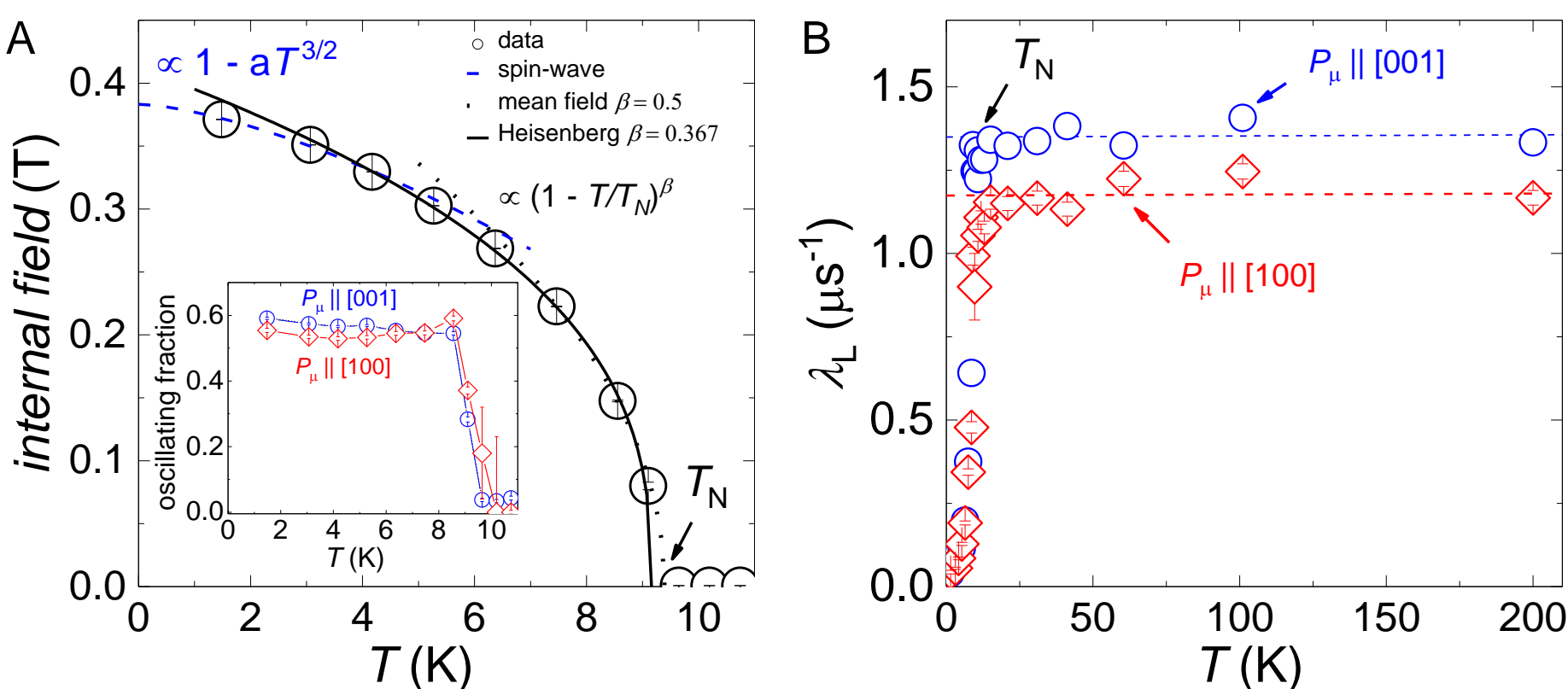

*Fig. 3. Temperature dependence of μSR. (A) Temperature dependence of the internal field obtained form zero field (ZF) μSR spectra below $T_N$. The overall behavior of the magnetic order parameter (~ internal field) is consistent with a 3D isotropic antiferromagnetic state. Inset: temperature dependnece of the amplitude of the oscilating singnal in ZF- μSR time spectra. (B) The temperature dependence of the longitudinal ZF relaxation rate measured for the muon spin polarisation components $P_\mu$ || [100] and || [001] directions. The data exclude any noticeable correlations effects above $T_N$.*

*Table 1. Charge carrier density, n; mobility, μ; negative magnetoresistance (nMR), anomalous Hall conductivity (AHC) and anomalous Hall angle (AHA) for GdPtBi and NdPtBi crystals at 2 K*

| Crystal | $n$ (cm$^{-3}$) ×10$^{18}$ | $\mu$ (cm$^2$V$^{-1}$s$^{-1}$) | nMR (%) | AHC (Ω$^{-1}$cm$^{-1}$) | AHA |
|---|---|---|---|---|---|
| GdPtBi-1 | 1.1 | 3360 | 63 | 60 | 0.23 |
| GdPtBi-2 | 2.7 | 3300 | 28 | 38 | 0.16 |
| NdPtBi | 0.8 | 1500 | 20 | 14 | 0.05 |

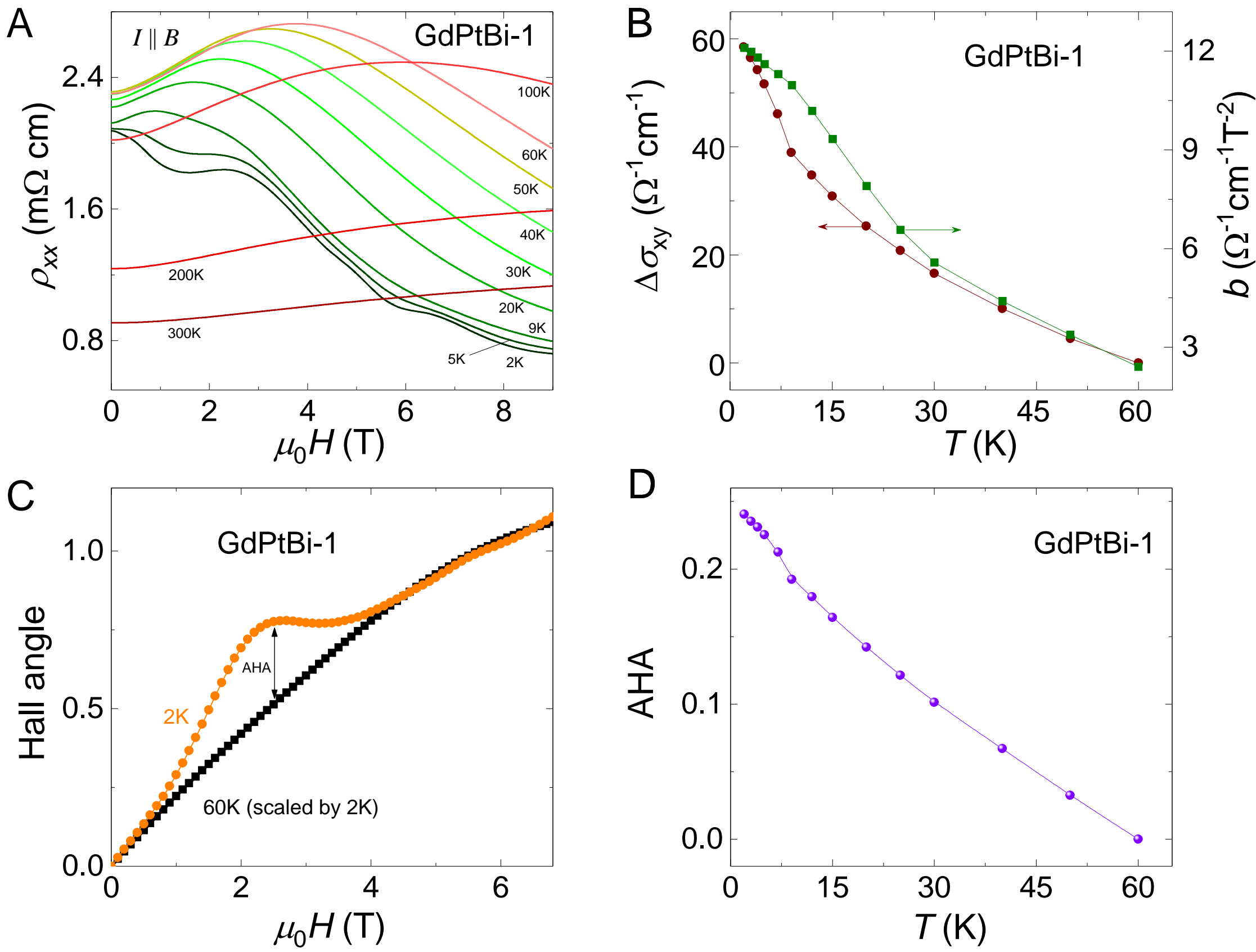

*Fig. 4. Negative magnetoresistance (nMR), anomalous Hall conductivity (AHC), $\Delta\sigma_{xy}$ and anomalous Hall angle (AHA), $\sigma_{xy}/\sigma_{xx}$. (A) nMR sustained up to 60 K in 9 T. (B) Temperature dependence AHC and coefficient of $B^2$, b which is estimated by fitting the data of A with $\sigma_{xx} = \sigma_0 + bB^2$, where $\sigma_{xx} = 1/\rho_{xx}$. (C) AHA at 2 K scaled with 60 K AHA data at 5 T and the subtracted AHA plotted against temperature in (D).*

Now we focus on the magnitude of the nMR. We investigate the evolution of the nMR with angle, $\theta$ as a function of temperature. Figure S9 shows the field dependence of $\rho_{xx}$ at various $\theta$ at 2 K. A large nMR is observed for $\theta \geq 60^{\circ}$ and a positive magnetoresistance appears which gradually increases as $\theta$ becomes less than 60°. For $\theta = 90^{\circ}$, the observed values of $\rho_{xx}$ at 0 and 9 T are 2.1 and 0.72 mΩ cm, respectively, that gives a MR of −66%. To explore the range of temperatures over which the nMR is found, we measured the field dependence of $\rho_{xx}$ at different temperatures for $\theta = 90^{\circ}$. As shown in Fig. 4A, $\rho_{xx}$ vs $B$ which can be divided into three regimes; 1) a moderate temperature dependence on temperature up to 12 K, 2) an increase of $\rho_{xx}$ with increasing the temperature up to 60 K, and 3) a positive magnetoresistance above ~60 K. A nMR below 100 K persists up to 33 T (above the saturation field of 25 T) as shown in Fig. S9. Remarkably, the highest nMR is observed at the lowest temperature and varies only smoothly across the magnetic transition. Such MR behavior rules out a magnetic cause and indicates the Weyl points as the origin. We have also simultaneously measured the magnetotransport properties of both NdPtBi and YPtBi. NdPtBi shows similar angular and temperature-dependent magnetotransport properties as GdPtBi. At $T = 2$ K NdPtBi exhibits a nMR of -20% for $I \parallel B$ in $B = 9$ T (Fig. S12). The MR decreases with increasing temperatures. In contrast to GdPtBi and NdPtBi, YPtBi does not show any AHE or nMR for similar temperature and field ranges (Fig. 2G-I) and Fig. S9B.

In an effort to find a relation between the AHE and nMR, we have used the function $\sigma_{xx} = \sigma_0 + bB^2$ to describe the longitudinal conductivity ($\sigma_{xx} = 1/\rho_{xx}$) over a wide range of temperature and field (Fig. S10, data of Fig. 4A). In this way, we obtained the coefficient, $b$ that quantifies the extent of the nMR. The temperature-dependent values of $\Delta\sigma_{xy}$ (detailed extraction method is mentioned in Fig. S8) describing the AHE and $b$ describing the nMR are shown in Fig. 4B. An excellent correlation between the magnitudes of $b$ and $\Delta\sigma_{xy}$ is observed over the entire temperature range (Fig. 4B). Thus, we conclude that AHE and nMR follow a similar trend with temperature, indicating their common origin. To extract the anomalous Hall angle (AHA) from the anomalous contribution of the Hall effect, we first plot the Hall angle as a function of field at various temperatures (Fig. 4C for data at 2K and Fig. S8 for other temperatures) and then scale them by the Hall angle data of 60 K (we assumed that there is no AHE at this temperature). The Hall angle at 2 K data exhibits a broad hump around 2 T while the Hall angle at 60 K data does not show such a hump. The difference between these two data sets gives an AHA value of 0.23 which is among the largest values reported so far. By applying this analysis to the Hall-angle data at different temperatures, the temperature-dependent AHA is obtained, as shown in Fig. 4D. The field dependent AHC at a particular temperature is calculated from the AHA by multiplying by the corresponding field dependent $\sigma_{xx}$ (Fig. S8C). Our investigation of the AHE and nMR which persist well above the $T_N$, establishes a strong correlation between both quantities. In GdPtBi, for which the Weyl points lie closer to the $E_F$, a large nMR together with a large AHE is found. Our *ab-initio* calculations predict a separation of the Weyl points along the [111] axis is as large as 30% of the Brillouin zone width, guaranteeing an observable AHE and chiral-anomaly-induced nMR. From the magnetotransport and magnetization data we can draw a phase diagram (Fig. S13) that shows how the various physical properties of GdPtBi depend on temperature and applied field. We sketch the regions for the nMR, positive MR and AHE from our field- and temperature-dependent measurements.

Our studies show that GdPtBi and NdPtBi become Weyl semimetals when the exchange splitting of the $\Gamma_8$ and $\Gamma_6$ bands is sufficiently large to establish the Weyl nodes. This, we find, happens even for small applied fields. It is clear from our experiments that this is not caused by simple external-field-induced Zeeman splitting but rather that the external field results in a significant alignment of the magnetization of the antiferromagnetic structure resulting in a large exchange field. Although the magnitude of this exchange field will increase up to the saturation field (~25 T at $T$ = 1.4 K), it is clear that once the exchange field is sufficiently large to establish the Weyl modes, further increases in the exchange field do not much affect the Weyl properties. We propose that the dip in resistivity and the corresponding peak in the Hall conductivity at about 2 T reflect this critical exchange value and the corresponding band-gap opening. We speculate that all magnetic rare-earth RPtBi and RAuSn (R=Ce-Sm, Gd-Tm) compounds will show related phase diagrams. For the systems CePtBi and YbPtBi the situation might be more complicated.

**Methods**

***Crystal growth:*** Single crystals of Y(Gd, Nd)PtBi were grown by a solution-growth method from a Bi flux. Freshly polished pieces of Y or (Gd or Nd), Pt and Bi, each of purity >99.99%, in the stoichiometric ratio (with significant excess Bi, i.e. Gd:Pt:Bi = 1:1:9) were placed in a tantalum crucible and sealed in a dry quartz ampoule under 3 mbar partial pressure of argon. The ampoule was heated at a rate of 100 K/h up to 1473 K and left for 12 hours at this temperature. For the crystal growth, the temperature was slowly reduced at a rate of 2 K/h to 873 K and the extra Bi flux was removed by decanting from the ampoule at 873 K. Typically we could obtain crystals, 1-5 mm in size, with a preferred growth orientation along [111], as confirmed by Laue diffraction. The methods we used follow closely those described in ref (29). The composition and structure was checked by energy dispersive X-ray analysis and Laue X-ray diffraction (Fig. S5), respectively. The lattice parameters of the cubic structure are 6.65 Å for YPtBi, 6.68 Å for GdPtBi and 6.76 Å for NdPtBi which are consistent with previous reports (29-31).

***Magnetoresistance and heat-capacity measurements***: Resistivity measurements were performed in a physical property measurement system (PPMS-9T, Quantum Design,) using the AC transport with rotator option. Heat capacity was measured by a relaxation method (HC option, PPMS, Quantum Design). Samples with bar shape of different crystalline orientations were cut from large single crystals using a wire saw. The orientation of these crystals was verified by Laue X-ray diffraction measurements and their physical dimensions are (width×thickness×length: $w \times d \times l$) 0.57×0.15×2.0 $mm^3$ for GdPtBi, 0.63×0.24×1.21 $mm^3$ for NdPtBi and 0.85×0.21×1.6 $mm^3$ for YPtBi. The linear contacts were made on the orientated crystals by silver paint and 25 μm platinum wires. The resistivity ($\rho_{xx}$) and Hall resistivity ($\rho_{yx}$) were measured in 4-wires and 5-wires geometry, respectively using a current of 1.0 mA at temperatures between 2 and 300 K and magnetic fields up to 9 T. Special attention was paid to the mounting of the samples on the rotating puck to ensure a good parallel alignment of the current and magnetic-field direction. The Hall resistivity contributions to the longitudinal resistivity and vice versa, due to contact misalignment, were accounted for by calculating the mean resistivity of positive and negative magnetic fields. Almost symmetrical longitudinal resistivities were obtained for positive and negative magnetic fields when current and magnetic field were parallel showing the excellent crystal and contact alignment of our samples. Otherwise, the nMR was overwhelmed by the transverse resistivities, i.e. Hall resistivity and trivial positive MR.

***μSR measurements:*** $\mu$SR experiments were performed on oriented GdPtBi single crystals at the GPS instrument of the $\pi$M3 beamline at the Paul Scherrer Institute (PSI) in Villigen, Switzerland. Fully spin-polarized, positive muons with a kinetic energy of 4.2 MeV were implanted in the sample (parallel to the crystallographic *a* axis [100] ) where they rapidly thermalize and stop at interstitial lattice sites at a depth of the order of 100 $\mu$m. For the $\mu$SR experiments we used two single crystals with the thickness ~ 0.5 mm placed side by side to increase the sample area (~ 15 mm$^2$) perpendicular to the muon beam. Using veto detectors allowed us to obtain the positron signal with a background contribution below 10% of the total signal. The measurements were performed in transversal polarization mode in which the muon spin polarization $P_\mu$ is at 45° with respect to the muon beam (pointing toward the upward counter) and sample *a* axis. The data were analyzed using the musrfit software package (32).

***Density-functional calculations:*** Density-functional theory (DFT) calculations were performed using the Vienna *ab-initio* Simulation Package (VASP) (33) with the generalized gradient approximation (GGA) (34) of the exchange-correlation energy. The experimentally measured lattice parameters were adopted and spin-orbit coupling was included in all calculations. 10×10×10 *k*-meshes were used for the Brillouin-zone sampling. An on-site Coulomb interaction was considered for the *f* electrons of Gd with $U$ = 10 eV. The band structures were projected to maximally localized Wannier functions (35). The chirality of the Weyl points was confirmed by identifying the source- or sink-type distribution of the Berry curvature.

**Acknowledgements:**

This work was financially supported by the ERC Advanced Grant No. 742068 'TOPMAT'. Part of the work was supported by DFG through the grant GR 4667 (V.G.) and through the SFB 1143. We also acknowledge support from the Dresden High Magnetic Field Laboratory (HLD) at HZDR and the High Field Magnet Laboratory (HFML-RU/FOM), members of the European Magnetic Field Laboratory (EMFL). $\mu$SR experiments were performed at Swiss Muon Source (SμS), PSI, Villigen.

**Author Contributions:** C.S. grew the crystals. C.S., N.K., S.S. and W.S. performed the transport and magnetic measurements. V.G., R.S., and H.L. performed $\mu$SR experiments. S.S, E.K., Y.S. J.W., A.MC. and U.Z. performed high-magnetic-field measurements. S.-C.W., Y.Z., B.Y. and J. K. carried out theoretical calculations. A.C.K. and C.S. performed the structural characterization. C.S., N.K., V.G., S.S., R.S., W.S., B.Y., H.H.K. S.S.P.P. and C.F. wrote the manuscript with inputs from all the authors. C.F. and S.S.P.P. jointly supervised the project.

# Supplementary Information Appendix

## A. Band structure and Weyl points

***Magnetic moments along [111].*** To unravel the emergence of the Weyl points, we introduce spin orbit coupling (SOC) and magnetism step by step in GdPtBi. As shown in Fig. S1, the band structure of GdPtBi without any magnetism and SOC exhibits $\Gamma_8$ and $\Gamma_6$ band inversion and is semimetallic. The non-magnetic cousin to GdPtBi, namely YPtBi, has similar characteristics. When magnetism is introduced by including Gd-4*f* electrons in the calculation, and considering the case when all the magnetic moments are fully aligned with each other, the energy bands are split due to the strong exchange coupling. The strength of the exchange field can be estimated through the band splitting at the Γ point, which is of the order of 0.5 eV (Fig. S2). The conduction and valence bands from different spin channels cross each other near the Fermi energy ($E_F$), forming a nodal-line like Fermi surface. When SOC is added, the band degeneracy of the nodal lines is lifted, except at some special points, i.e. the Weyl points. We note that the direction of the magnetic moments is very sensitive upon tuning the positions of the Weyl points as well as the Fermi surface topology, as we will discuss in the following.

When the Gd magnetic moments are oriented along the [111], the body diagonal of the *fcc* unit cell, the system exhibits $C_3$ rotational symmetry with respect to the [111]. In the band structure, one can clearly observe the Weyl points along the Γ – L [111], as a crossing point between the valence band and the conduction band, which lies 39 meV below the Fermi energy (Fig. S2A). Along the opposite [-1-1-1], there is another Weyl point with the opposite chirality at the same energy. We label this pair of Weyl points along the [111] as W1. Moreover, we have found six additional Weyl points away from the high-symmetry lines, as shown in Fig. S2B. Those with the same chirality lie in a (111) plane and can be transformed into each other by a $C_3$ rotation, since rotation does not violate the chirality. We label these three pairs of Weyl points as W2.

We list the energies, chiralities and positions of all eight Weyl points in Fig. S2C. Here, we identify the chirality by calculating the Berry curvature near the Weyl points, where a sink of Berry curvature is named "–" and a source "+". The monopole charges (Chern numbers) of the Weyl points are found to be ± 1 from the integral of the Berry curvature over a closed sphere that encloses a Weyl point.

***Magnetic moments along [001].*** When the Gd moments point along [001], the system is reduced to the $S_4$ symmetry, i.e. a $C_4$ rotation around the [001] followed by a mirror reflection. In the band structure, one then finds that the band crossing is gapped along the high-symmetry lines (Fig. S3). We find six pairs of Weyl points away from high symmetry lines and planes: four pairs below the Fermi energy (labelled W1) and two pairs above the Fermi energy (labelled W2). The four W1 Weyl points with the same chirality are located in the (001) plane. They can be transformed to the other four W1 points which have the opposite chirality by a $S_4$ operation. The W2 Weyl points appear 48 meV above the Fermi energy. The "±" W2 points are also connected to each other by the $S_4$ symmetry transformation.

***Magnetic moments along [110].*** We should point out that the Weyl points found for the [111] and [001] cases are all type-I Weyl points, i.e. the Weyl cones are in the typical hourglass shape. However, we have observed type-II Weyl points with strongly tilted Weyl cones when the moments are oriented along the [110] direction. As shown in Fig. S4, two pairs of Weyl points exist, where a pair of Weyl points with the opposite chirality is connected by a reflection with respect to the (110) plane.

## Single-crystal X-ray diffraction.

The crystal structure of GdPtBi has been determined by means of single crystal X-ray diffraction using a *Bruker D8 VENTURE* X-ray diffractometer equipped with a bent graphite monochromator for Mo $K\alpha$ radiation and a *Photon* large area detector. 2465 reflections have been collected up to $2\Theta_{max} = 116.3°$. All reflections could be indexed on the basis of a face-centered cubic Bravais lattice with $a$ = 6.6933(4) Å. Fig. S5 shows the X-ray scattering intensities in the $hk0$, $0kl$ and $h0l$ planes of reciprocal space that indicate the single crystallinity of our sample. In accordance with previous reports in literature, a structural refinement with space group No. 216, $F\bar{4}3m$ yields a proper crystal structure refinement with $R$ and $R_w$ values that amount to 1.83% and 4.53% respectively.

Also all obtained thermal displacement parameters are reasonably small, thus, indicating no substantial disorder within our single crystals, see Supplementary table S1.

***Structural parameters.*** Atomic coordinates $x, y, z$ and thermal displacement parameters $U_{ii}$ for GdPtBi. The crystal structure refinement has been performed by using the *Jana2006* software

Supplementary table S1

| Atom | $x$ | $y$ | $z$ | $U_{ii}$ |
|---|---|---|---|---|
| Bi | 0 | 0 | 0 | 0.00446(6) Å$^2$ |
| Pt | 1/4 | ¼ | 1/4 | 0.00601(5) Å$^2$ |
| Gd | 1/2 | ½ | 1/2 | 0.00449(9) Å$^2$ |

**Resistivity, mobility and charge-carrier density of RPtBi.**

The RPtBi compounds, where R is a rare-earth element or Y, are members of the half-Heusler family of compounds. The temperature dependence of the resistivity, mobility and charge-carrier density of NdPtBi and GdPtBi and YPtBi are shown in Fig. S6. We observe a small feature in the temperature dependence of the resistivity in NdPtBi and GdPtBi compounds at the same temperature as their antiferromagnetic transition temperatures: i.e. at $T_N$ = 2.1 K and 9.0 K for NdPtBi and GdPtBi, respectively. However, there is no such feature for the nonmagnetic YPtBi. Using a single-band model, the relation between the Hall coefficient and the field and resistivity are $R_H = \rho_{xy}/B$ and $R_H = 1/ne$, where $R_H$ is Hall coefficient, $\rho_{xy}$ is Hall resistivity, $B$ is applied field, $n$ is carrier density and $e$ is the elementary charge. We calculate the mobility and charge-carrier density of all three compounds from the above relations wherein $R_H$ is estimated by taking the slope of the temperature dependent $\rho_{xy}$ for fields within the range 4-9 T. The majority charge carriers are holes for all three compounds over the entire temperature range. At $T$ = 2 K, the mobility and charge-carrier density are of the order of $10^3$ cm$^2$/Vs and $10^{18}$ cm$^{-3}$, respectively.

**Magnetization and specific heat.**

Figure S7A displays the temperature dependence of the susceptibility ($M/H$) as a function of temperature for an applied magnetic field of 0.1 T. A clear AFM transition is observed at $T_N$ = 8.8 K. The inverse susceptibility ($H/M$) can be described by use of the Curie- Weiss law

$H/M = (T - \theta_W)/C$, where $\theta_W$ is the Curie- Weiss temperature and $C$ is the Curie constant, which was determined from the slope of $H/M$. The derived effective moment from the $C$ is 7.78 $\mu_B$. This is close to the moment of a free $Gd^{3+}$ ion. The value $\theta_W$ = 33 K is in well agreement with earlier report (1). Fig. S7B shows the isothermal magnetization curve at $T$ = 2 K taken up to 7 T. This shows almost linear behavior.

The heat capacity of GdPtBi was measured by a relaxation method (HC option, PPMS, Quantum Design) on a single crystal ($m$ = 5.538 mg). The specific heat, $c_p(T)$, of GdPtBi (Fig. S7C) shows a typical lambda-type peak due to long-range order of Gd magnetic moments at the $T_N$ = 8.6 K. With increasing magnetic field $T_N$ decreases ($T_N$ = 8.1 K for $\mu_0 H$ = 9 T), indicating an antiferromagnetic ordered structure (1). For analysis we assume that $C_p(T)$ is the sum of phonon and magnetic contributions, thus $C_{tot}(T) = C_{lat}(T) + C_{mag}(T)$, neglecting a Sommerfeld contribution. No reference sample for $C_{lat}(T)$ was at hand, therefore, a method based on equivalent Debye temperature (2) was adopted to estimate $C_{lat}(T)$. Since for Gd with an $^8S_{7/2}$ ground state, the magnetic order involves the complete multiplet, there is besides short-range order contributions no further magnetic specific heat above $T_N$.

**Magneto-transport.**

Figure S8 shows the field-dependent $\rho_{xx}$ at various angles, $\theta$, between the applied field and the current direction at 2 K. A large negative magnetoresistance (nMR) is observed for $\theta \geq 60^o$ and a positive magnetoresistance appears which gradually increases as $\theta$ decreases from 60°. We calculate the magnetoresistance MR by using the relation, MR (%) = $[\rho_{xx}(B) - \rho_{xx}(0)]/\rho_{xx}(0) \times 100$ yeilding -66 % at 2 K and 9 T when I || B. The MR remains negative up to 33T.

NdPtBi is isostructural to GdPtBi and shows an anomalous Hall conductivity (AHC) of 14 $\Omega^{-1}cm^{-1}$ and a MR of -20% at 2 K in 9 T as shown in Fig. S11 and S12, respectively . The MR is still negative above 50° and temperatures up to 10 K at 90°. From the magneto-transport and magnetization measurements, we construct a phase diagram shown in Fig. S13 that reveals how the various physical properties of GdPtBi depend on temperature and applied field. We sketch the regions for the nMR, positive MR and AHE from the field and temperature-dependent measurements.

### ZF μSR time spectra

To fit the ZF asymmetry time spectra we used:

$$A(t) = A_s(0)[\alpha \cos(2\pi \nu t+\varphi) \exp(-\lambda_T t)+(1-\alpha) \exp(-\lambda_L t)] + A_{bg}; \qquad \text{(S1)}$$

where $A_s(0)$ is the initial asymmetry of the sample signal, $A_{bg}$ is the contribution of the background to the total signal (in the analysis we assumed that $A_{bg}$ is temperature and time independent), $\nu = \gamma_\mu B_{int}$ is the frequency of the asymmetry oscillations due to static internal magnetic field $B_{int}$ with $\gamma_\mu$ = 135.5342 MHz $T^{-1}$, α is a fraction of the signal with oscillations, $\lambda_T$ is the transversal relaxation rate, and $\lambda_L$ is the longitudinal relaxation rate, which is related to magnetic fluctuations (see Fig. S14).

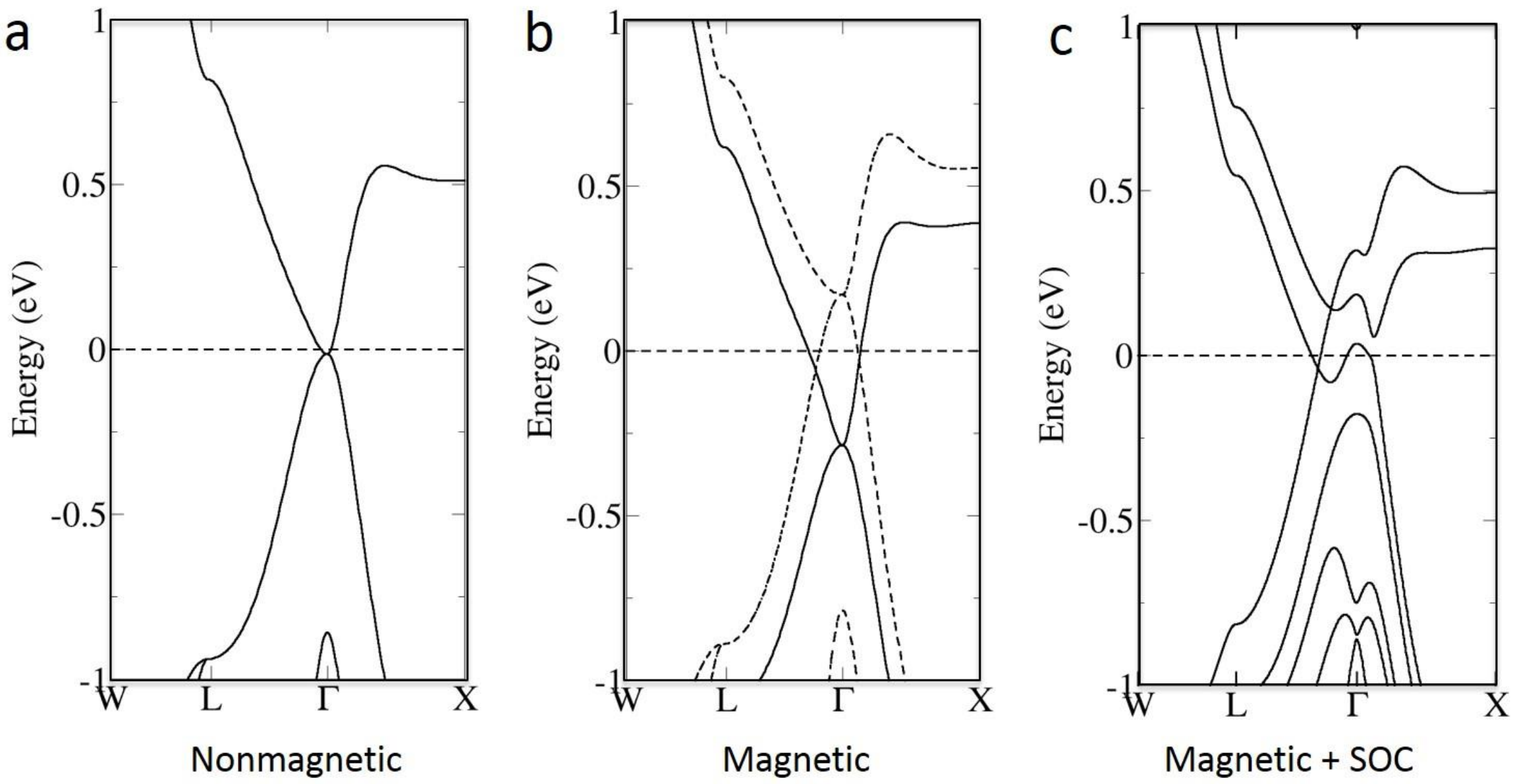

***Fig. S1.*** *Band structure of GdPtBi. (**A**) The non-magnetic case without including Gd-f electrons and spin-orbit coupling (SOC). (**B**) The magnetic case without SOC. (**C**) Inclusion of magnetism and SOC when all the magnetic moments are fully oriented along [111].*

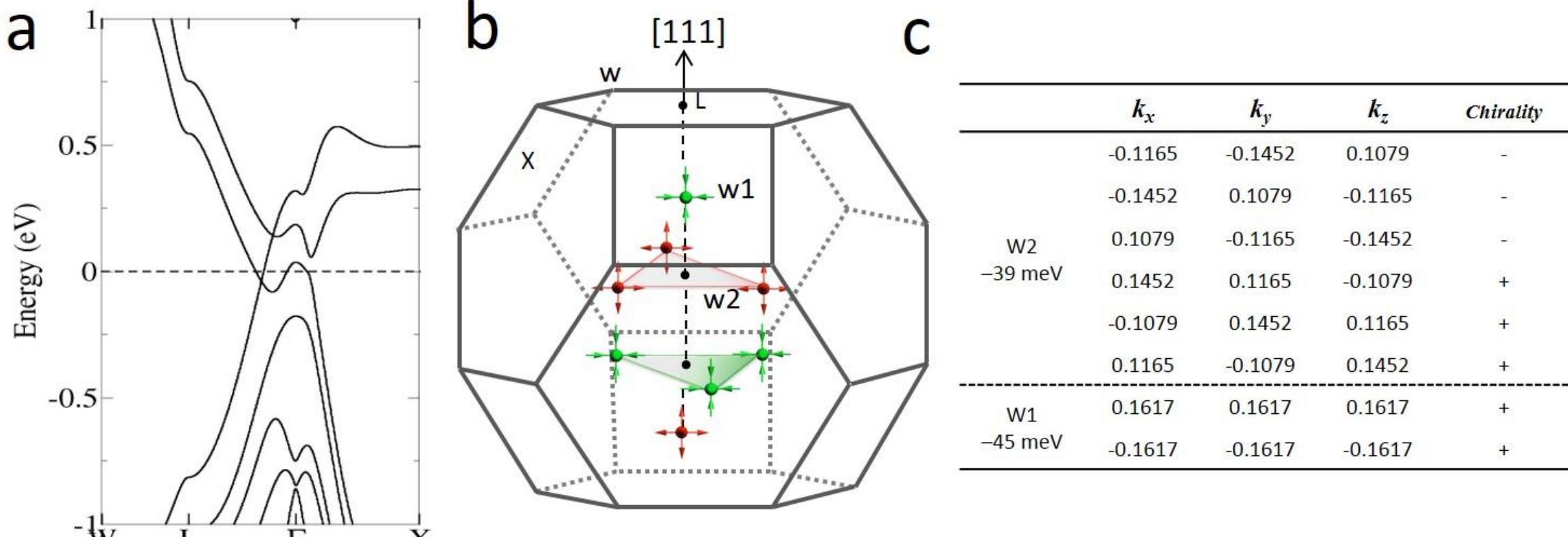

| | $k_x$ | $k_y$ | $k_z$ | Chirality |
|---|---|---|---|---|
| W2 –39 meV | -0.1165 | -0.1452 | 0.1079 | - |
| | -0.1452 | 0.1079 | -0.1165 | - |
| | 0.1079 | -0.1165 | -0.1452 | - |
| | 0.1452 | 0.1165 | -0.1079 | + |
| | -0.1079 | 0.1452 | 0.1165 | + |
| | 0.1165 | -0.1079 | 0.1452 | + |
| W1 –45 meV | 0.1617 | 0.1617 | 0.1617 | + |
| | -0.1617 | -0.1617 | -0.1617 | + |

***Fig. S2.*** *The band structure and Weyl points when the Gd moments are oriented along [111].* ***(A)*** *The band structure along high-symmetry lines. The Fermi energy is shifted to zero.* ***(B)*** *The first Brillouin zone and four pairs of Weyl points. Red and green spheres represent Weyl points with "+" and "–" chirality, respectively. The arrows represent the Berry curvature near the Weyl points.* ***(C)*** *The energies, positions and chiralities of all Weyl points. The positions are in unit of 2π/a, where a = 6.68 Å is the lattice parameter.*

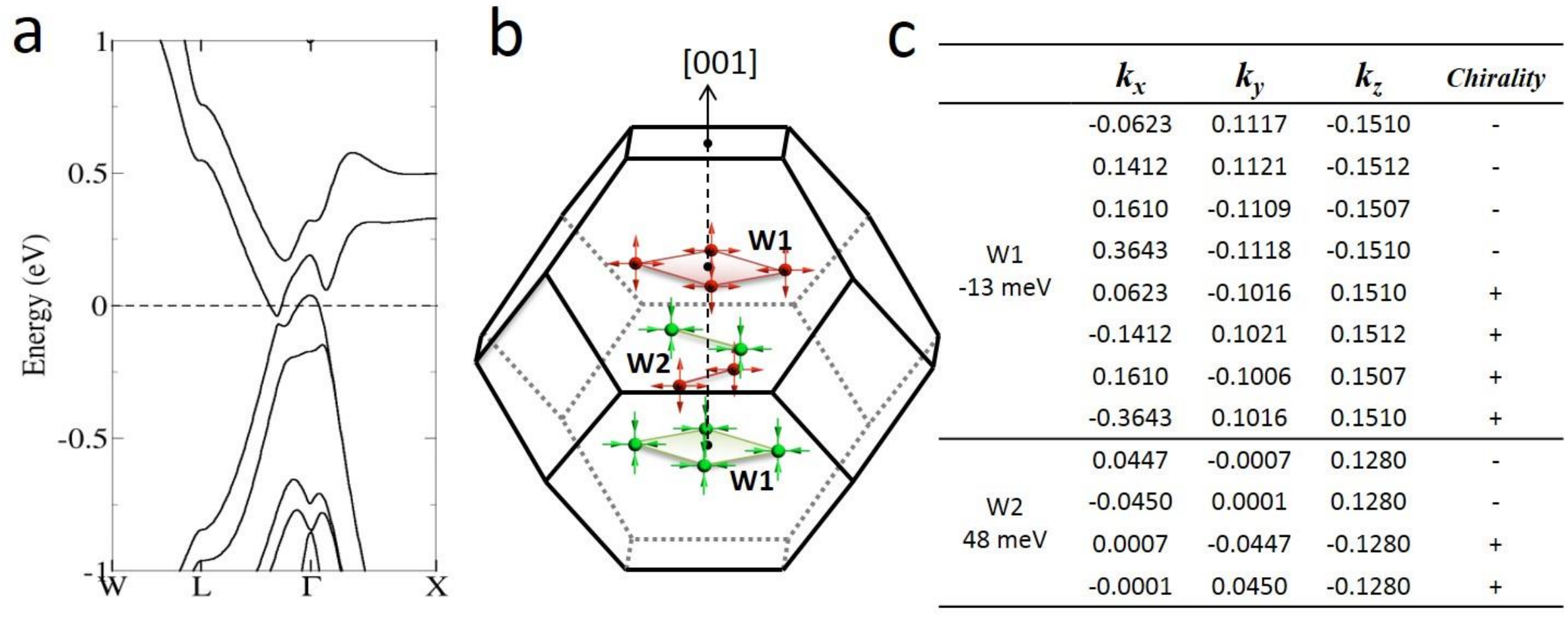

| | $k_x$ | $k_y$ | $k_z$ | Chirality |
|---|---|---|---|---|
| W1 -13 meV | -0.0623 | 0.1117 | -0.1510 | - |
| | 0.1412 | 0.1121 | -0.1512 | - |
| | 0.1610 | -0.1109 | -0.1507 | - |
| | 0.3643 | -0.1118 | -0.1510 | - |
| | 0.0623 | -0.1016 | 0.1510 | + |
| | -0.1412 | 0.1021 | 0.1512 | + |
| | 0.1610 | -0.1006 | 0.1507 | + |
| | -0.3643 | 0.1016 | 0.1510 | + |
| W2 48 meV | 0.0447 | -0.0007 | 0.1280 | - |
| | -0.0450 | 0.0001 | 0.1280 | - |
| | 0.0007 | -0.0447 | -0.1280 | + |
| | -0.0001 | 0.0450 | -0.1280 | + |

***Fig. S3.*** *The band structure and Weyl points for moments along [001].* ***(A)*** *The band structure along high-symmetry lines.* ***(B)*** *The first Brillouin zone and six pairs of Weyl points.* ***(C)*** *The energies, positions and chiralities of all Weyl points.*

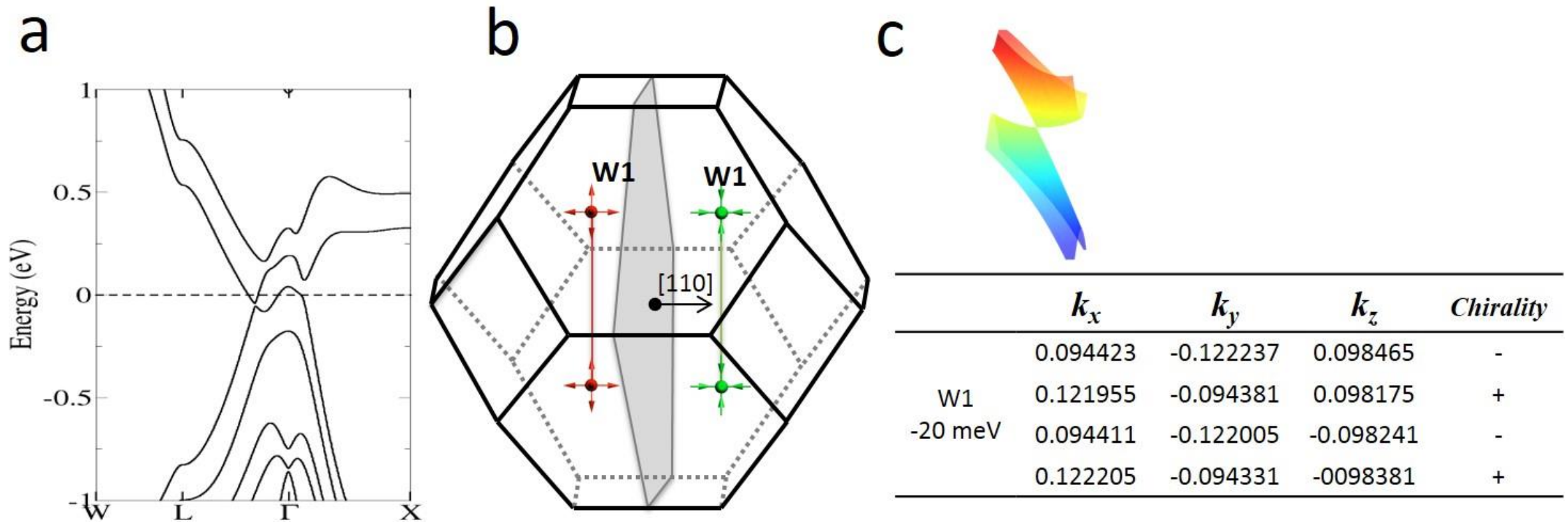

|  | $k_x$ | $k_y$ | $k_z$ | *Chirality* |
|---|---|---|---|---|
| W1<br>-20 meV | 0.094423 | -0.122237 | 0.098465 | - |
|  | 0.121955 | -0.094381 | 0.098175 | + |
|  | 0.094411 | -0.122005 | -0.098241 | - |
|  | 0.122205 | -0.094331 | -0098381 | + |

***Fig. S4.*** *The band structure and Weyl points for moments along [110].* ***(A)*** *The band structure along high-symmetry lines.* ***(B)*** *The first Brillouin zone and six pairs of Weyl points.* ***(C)*** *The energies, positions and chiralities of all Weyl points. Close-up of the Weyl cone near the Weyl point is shown to demonstrate the type-II tilting feature.*

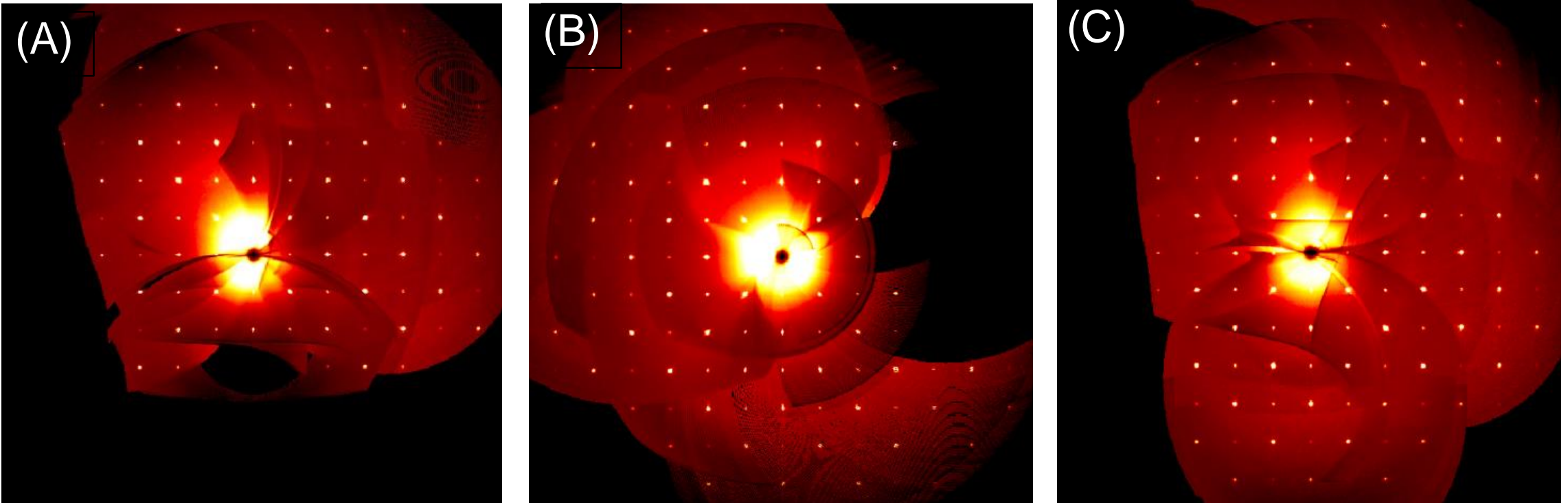

***Fig. S5.*** *Precession-maps X-ray scattering intensities of GdPtBi-1 observed within the;* ***(A)*** *hk0.* ***(B)*** *0kl.* ***(C)*** *h0l plane of the reciprocal space.*

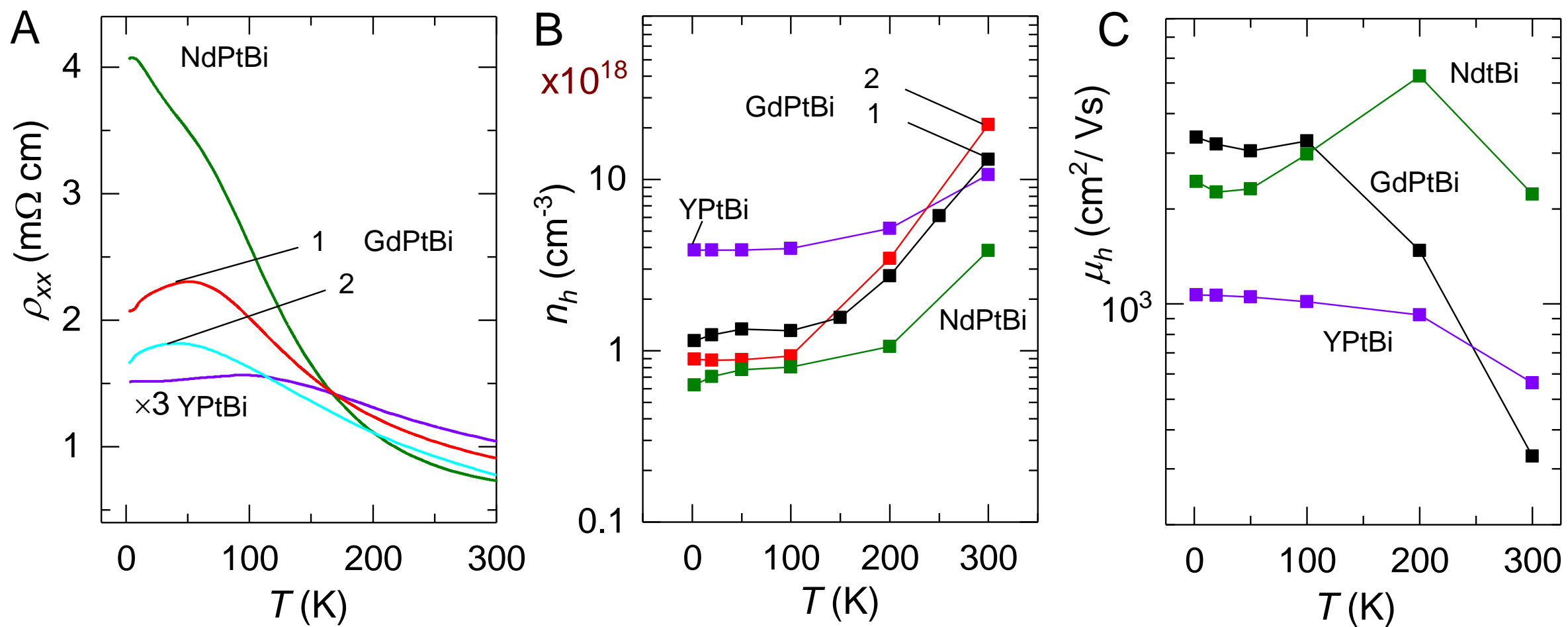

***Fig. S6.*** *Temperature-dependent resistivity, mobility, μ and charge-carrier density, n.* ***(A)*** *Comparison of the temperature-dependent resistivities for GdPtBi (2 crystals), NdPtBi and YPtBi. A clear feature of the antiferromagnetic transition is observed in GdPtBi and NdPtBi at* $T_N$ *= 9 K and 2.1 K, respectively.* ***(B)*** *μ and* ***(C)*** *n of all three compounds as a function of temperature.*

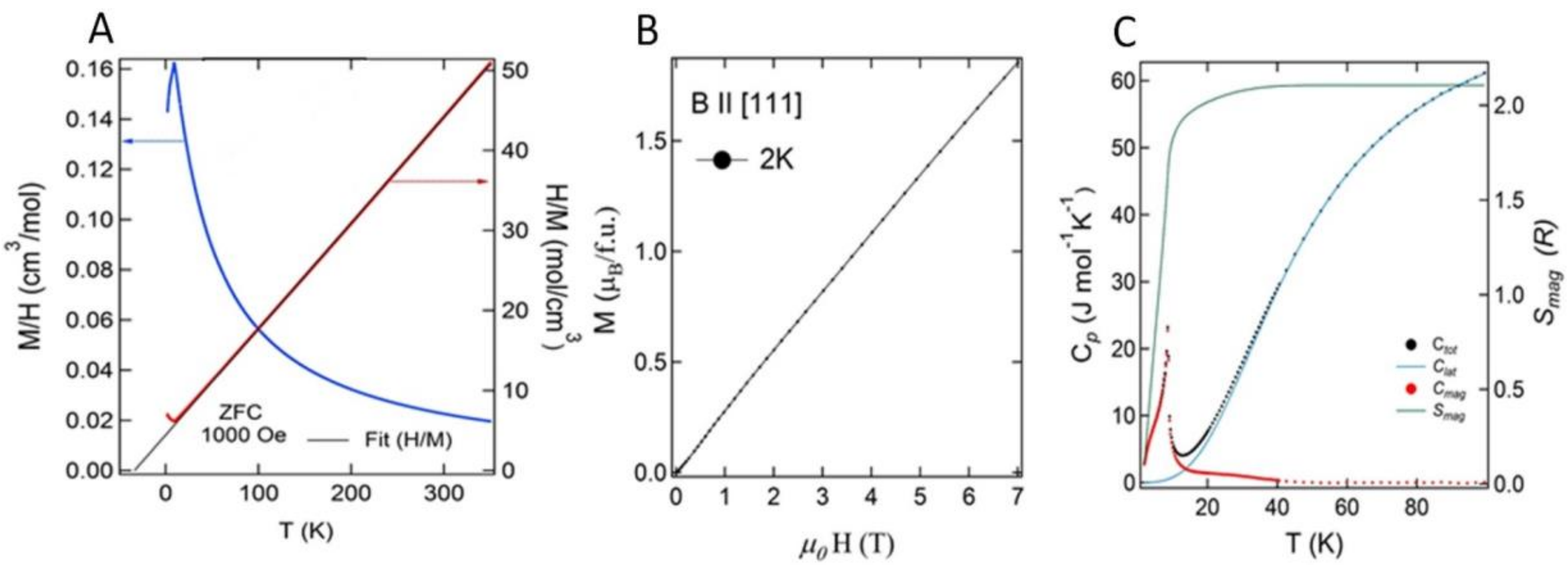

***Fig. S7.*** *Magnetization and specific heat of GdPtBi-1.* ***(A)*** *Temperature dependence of susceptibility (M/H) as a function of temperature for an applied magnetic field of 0.1 T (left axis). The antiferromagnetic transition at* $T_N$ *= 8.8 K is clearly seen. The inverse magnetic susceptibility (H/M) follows Curie-Weiss law.* ***(B)*** *Isotherm magnetization curve at 2 K for B || [111].* ***(C)*** *Temperature dependence of the molar capacity heat (*$C_{tot}$*), separated in to the lattice (*$C_{lat}$*), and magnetic contribution (*$C_{mag}$*). The magnetic entropy* $S_{mag}(T)$ *in units of R (molar gas constant* $R = N_A k_B$*) is obtained by integration of* $C_{mag}/T$ *and includes an extrapolation to T = 0.*

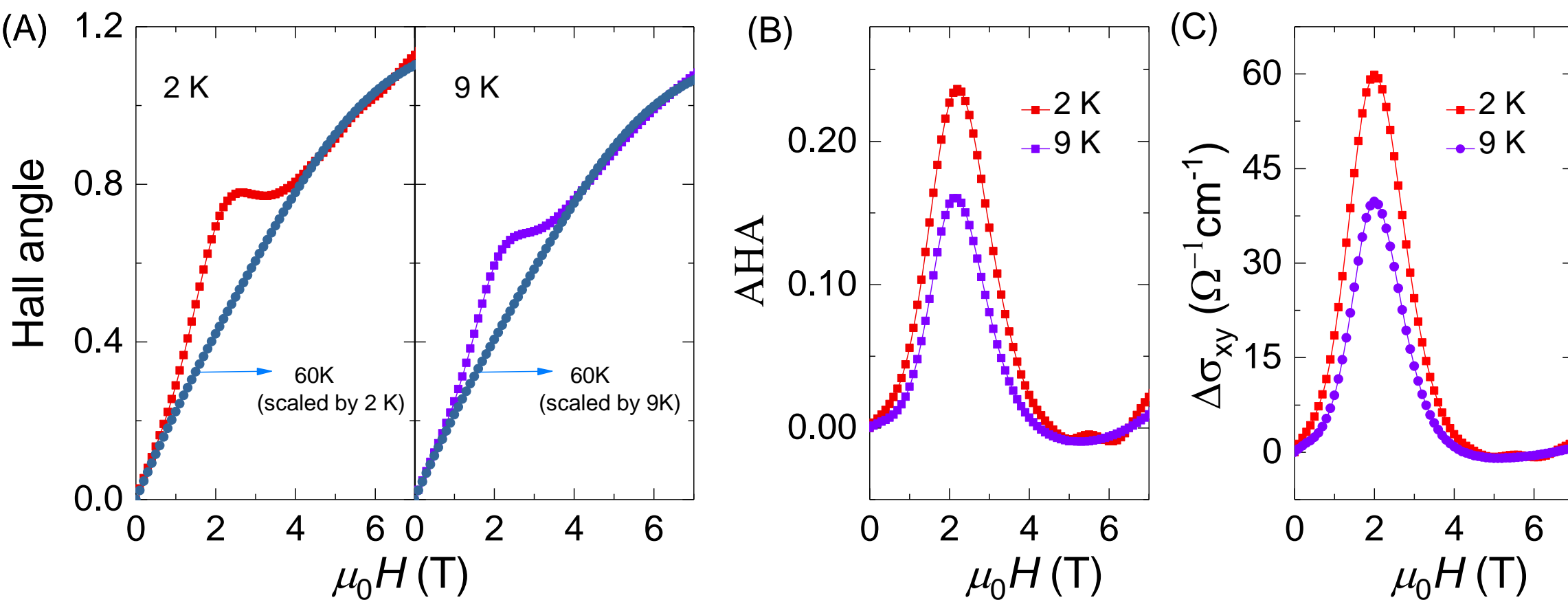

***Fig. S8.*** *Determination of temperature dependent anomalous Hall angle (AHA) and anomalous Hall conductivity (AHC),* $\Delta\sigma_{xy}$ *for GdPtBi-1. (A) Field dependent Hall angle is calculated at different temperature up to 60 K. (B) To get temperature dependent AHA, each temperature data set of Hall angle is subtracted by scaled the data of 60 K at 5 T, for example 2 and 9 K data are given here. (B) For AHC, AHA data of panel B are multiplied by respective value of* $\sigma_{xx}$. $\sigma_{xy}$ *and* $\sigma_{xx}$ *are calculated from the relation* $\sigma_{xy} = \frac{\rho_{yx}}{\rho_{yx}^2+\rho_{xx}^2}$ *and* $\sigma_{xx} = \frac{\rho_{xx}}{\rho_{yx}^2+\rho_{xx}^2}$, *respectively. Hall angle defines a ratio of the Hall conductivity,* $\sigma_{xy}$ *and the traverse conductivity,* $\sigma_{xx}$.

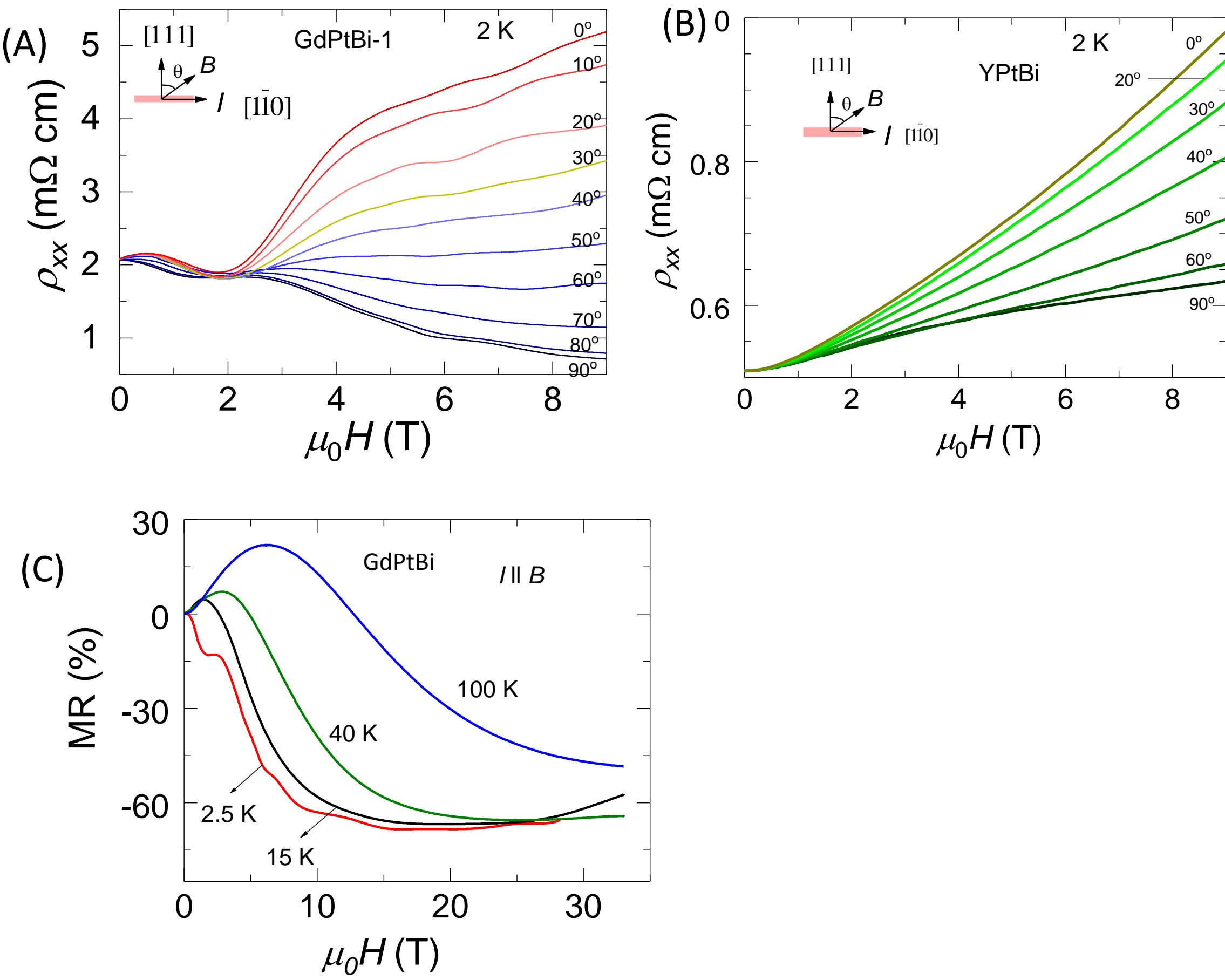

***Fig. S9.*** *Angular-dependent $\rho_{xx}$ of GdPtBi-1 and YPtBi. Field-dependent resistivity, $\rho_{xx}$ at various $\theta$ from $\theta = 0^o$ ($I \perp B$) to $\theta = 90^o$ ($I \,//\, B$) at $T = 2K$ for current along* [111] *(A) for GdPtBi-1 and (B) for YPtBi. Longitudinal MR ($I \,//\, B$) measured at different temperatures and fields up to 33 T for GdPtBi.*

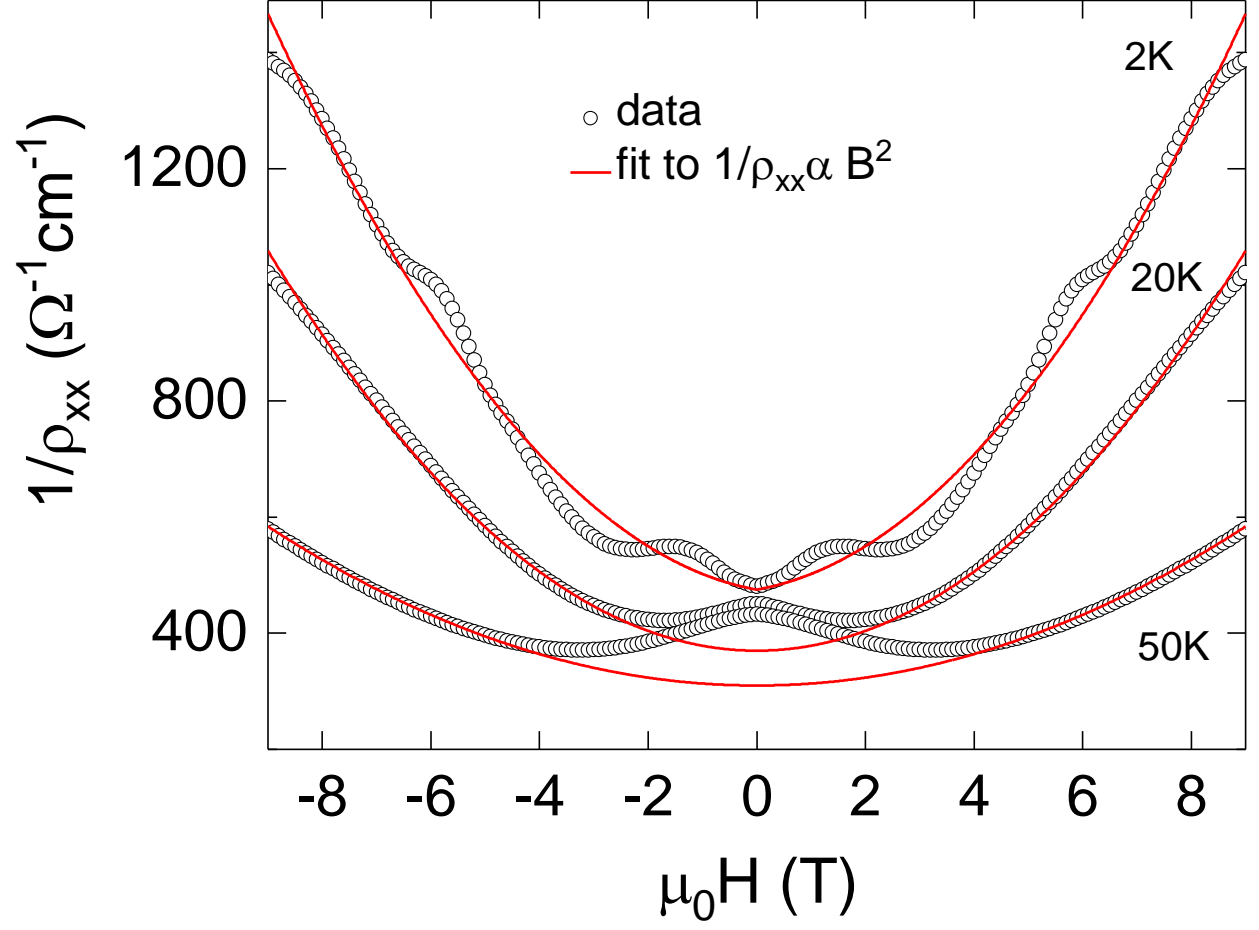

***Fig. S10.*** *Different fits (red lines) describing the field dependent longitudinal conductivity, $1/\rho_{xx}$, at various temperatures for GdPtBi-1. To obtain the coefficient of B, b, we fitted (red line) with* $1/\rho_{xx} = \sigma_{xx} = \sigma_0 + bB^2$.

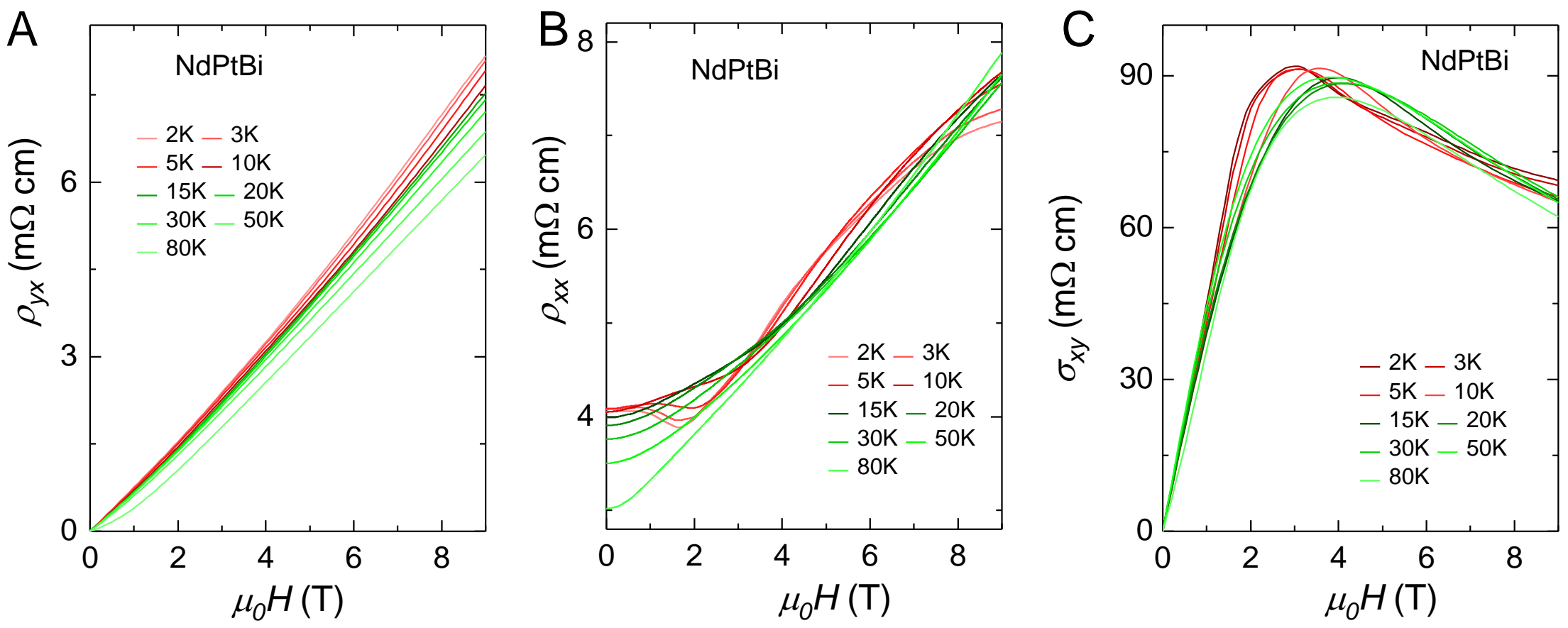

***Fig. S11.*** *$\rho_{xy}$, $\rho_{xx}$ and corresponding $\sigma_{xx}$ for NdPtBi. Field dependence of* ***(A)*** *$\rho_{xy}$.* ***(B)*** *$\rho_{xx}$ and.* ***(C)*** *$\sigma_{xy}$ at different temperatures.*

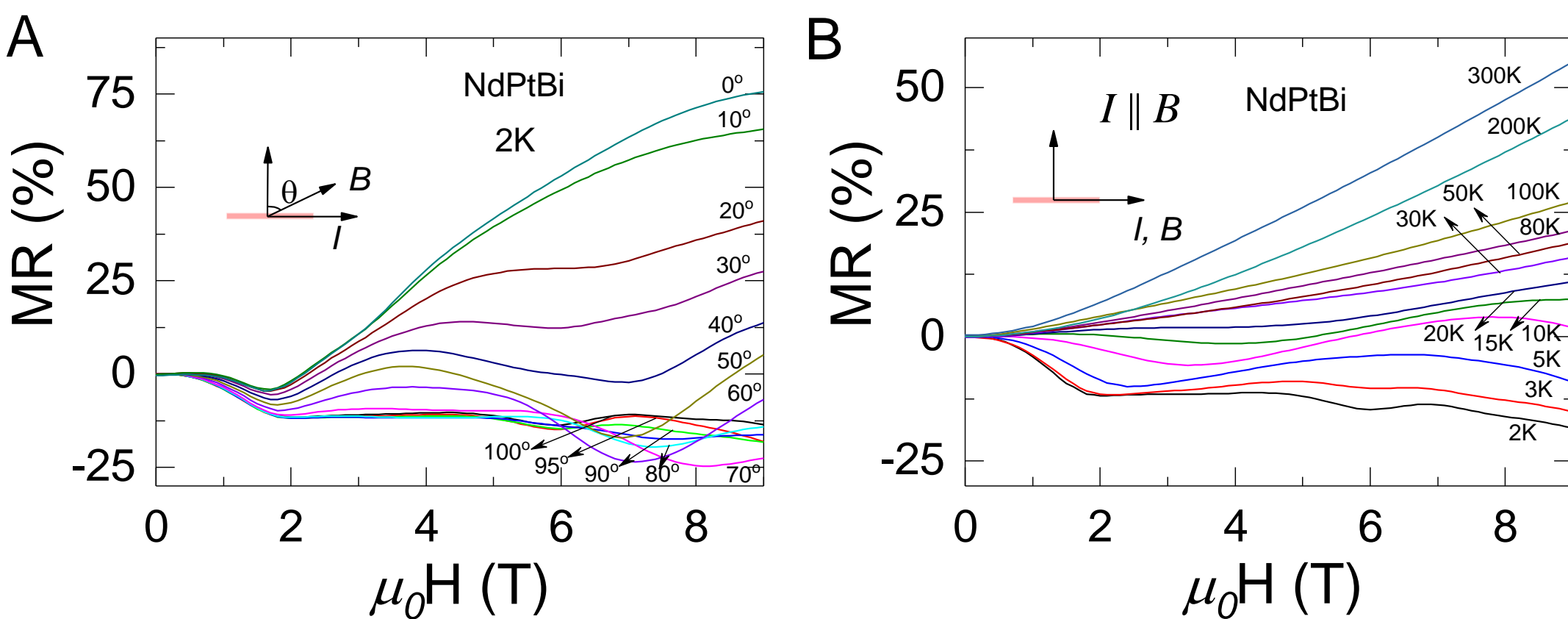

***Fig. S12.*** *Angular-dependence of the MR of NdPtBi.* ***(A)*** *MR of NdPtBi between 0$^o$ to 100$^o$ at 2 K.* ***(B)*** *shows longitudinal MR (I//B) of NdPtBi at different temperatures.*

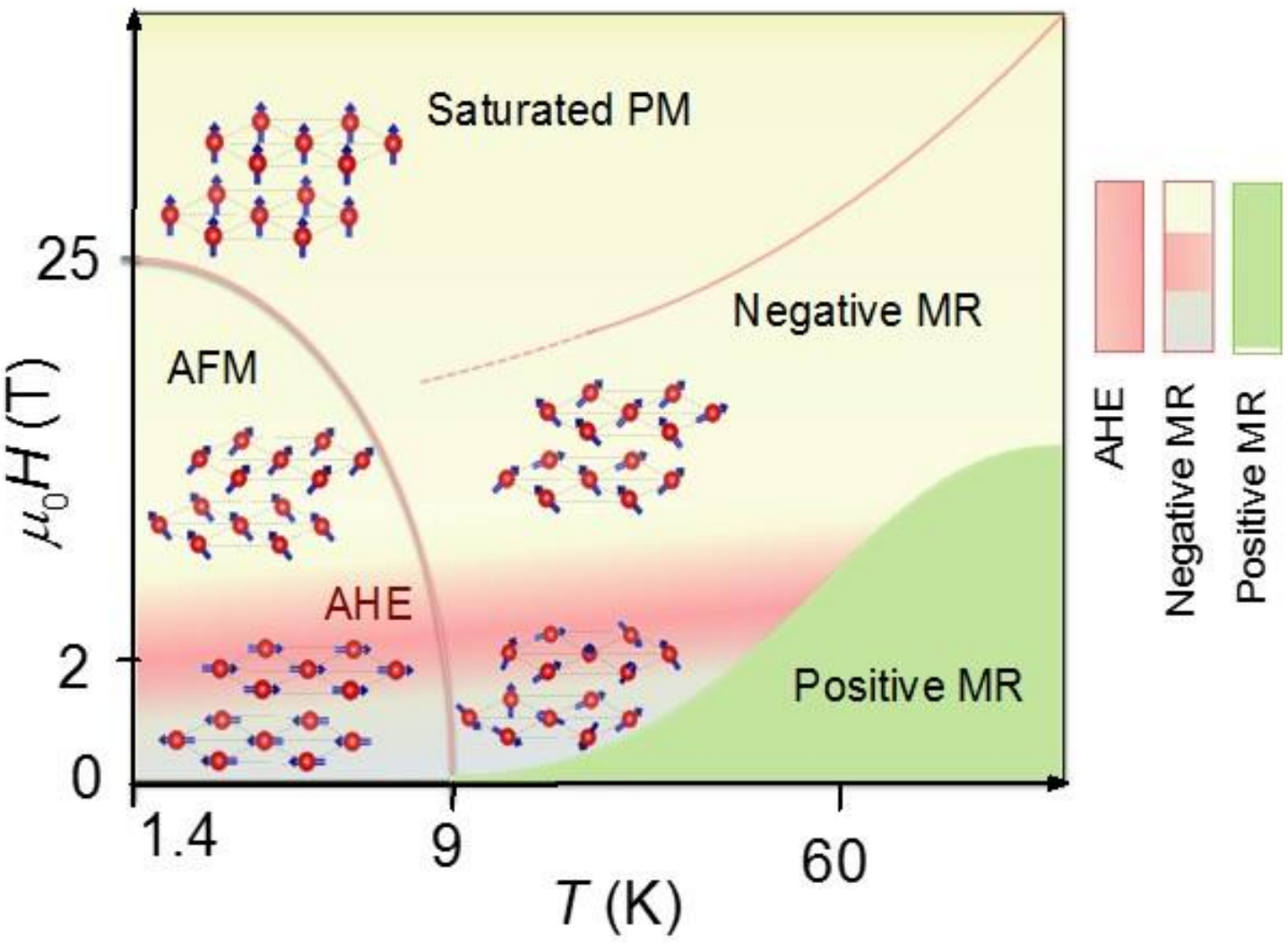

***Fig. S13.*** *Schematic H-T phase diagrams of GdPtBi. Below $T_N$ (9 K), the magnetic structure evolves from a type-II AFM state at small field to a fully saturated magnetic sate for higher fields. Around 2 T up to ~60 K, AHE is represented by pink shadow. For I// B, the negative MR persists up to very high magnetic field and temperatures while the positive MR appears only above $T_N$ and is limited to small field. The color codes for the different regions of the phase diagram are shown on the right. It should be noted that the AHE is for I⊥B while the other effects (negative MR and positive MR) are for I//B. The AHE and negative MR are observed over a wide temperature range that extends from below to above $T_N$.*

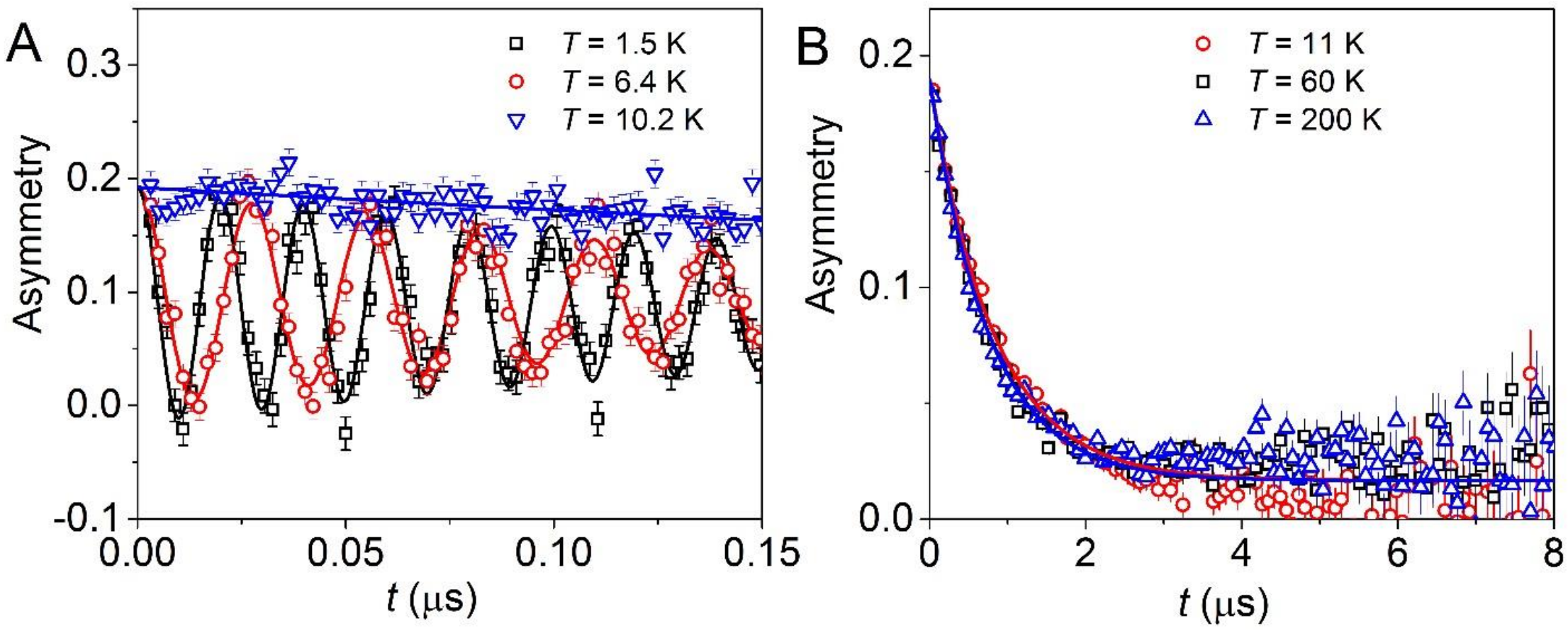

***Fig. S14.*** *Representative ZF-μSR asymmetry time spectra of GdPtBi: (A) above and below $T_N$, and (B) in the paramagnetic state. Symbols and solid lines are the experimental data and theoretical descriptions as detailed in the text (equation S1), respectively.*